\documentclass[aoas,MSNbibl,nameyear,dvips]{arximspdf}
\usepackage{dcolumn}
\usepackage{graphicx}

\doi{10.1214/12-AOAS564} 
\volume{6}
\issue{4}
\pubyear{2012}
\firstpage{1452}
\lastpage{1477}

\makeatletter

\newcolumntype{d}[1]{D{.}{.}{#1}}

\newcommand{\mat}[1]{{\bolds #1}}
\newcommand{\matt}[1]{{\mathbf #1}}

\newcommand{\vect}[1]{{\bolds #1}}

\makeatother

\begin{document}
\begin{frontmatter}

\title{A dynamic nonstationary spatio-temporal model for short term
prediction of precipitation}
\runtitle{A spatio-temporal model for precipitation}

\begin{aug}
\author[A]{\fnms{Fabio} \snm{Sigrist}\corref{}\ead[label=e1]{sigrist@stat.math.ethz.ch}},
\author[A]{\fnms{Hans R.} \snm{K\"unsch}\ead[label=e2]{kuensch@stat.math.ethz.ch}}
\and
\author[A]{\fnms{Werner A.} \snm{Stahel}\ead[label=e3]{stahel@stat.math.ethz.ch}}
\runauthor{F. Sigrist, H. R. K\"unsch and W. A. Stahel}
\affiliation{Seminar for Statistics, ETH Zurich}
\address[A]{Seminar for Statistics\\
Department of Mathematics\\
ETH Zurich\\
Switzerland\\
\printead{e1}\\
\hphantom{E-mail: }\printead*{e2}\\
\hphantom{E-mail: }\printead*{e3}}
\end{aug}

\received{\smonth{12} \syear{2011}}
\revised{\smonth{3} \syear{2012}}

%
\begin{abstract}
Precipitation is a complex physical process that varies in space and
time. Predictions and interpolations at unobserved times and/or
locations help to solve important problems in many areas. In this
paper, we present a hierarchical Bayesian model for spatio-temporal
data and apply it to obtain short term predictions of rainfall. The
model incorporates physical knowledge about the underlying processes
that determine rainfall, such as advection, diffusion and convection.
It is based on a temporal autoregressive convolution with spatially
colored and temporally white innovations. By linking the advection
parameter of the convolution kernel to an external wind vector, the
model is temporally nonstationary. Further, it allows for nonseparable
and anisotropic covariance structures. With the help of the Voronoi
tessellation, we construct a natural parametrization, that is, space as
well as time resolution consistent, for data lying on irregular grid
points. In the application, the statistical model combines forecasts of
three other meteorological variables obtained from a numerical weather
prediction model with past precipitation observations. The model is
then used to predict three-hourly precipitation over 24 hours. It
performs better than a separable, stationary and isotropic version, and
it performs comparably to a deterministic numerical weather prediction
model for precipitation and has the advantage that it quantifies
prediction uncertainty.
\end{abstract}

%

\end{frontmatter}

\section{Introduction}
Precipitation is a very complex phenomenon that varies in space and
time, and there are many efforts to model it.
Predictions and interpolations at unobserved times and/or
locations obtained from such
models help to solve important problems in areas such as
agriculture, climate science, ecology and hydrology. Stochastic models
have the great advantage of providing not only point estimates, but
also quantitative measures of uncertainty. They
can be used, for instance, as stochastic generators [\citet{Wi98},
\citet{MaMc09}] to provide
realistic inputs to flooding, runoff and crop growth models. Moreover,
they can be applied as components within general circulation models used
in climate change studies [\citet{FoEtAl05}], or for
postprocessing precipitation
forecasts [\citet{SlRaGn07}].

\subsection{Distributions for precipitation}
A characteristic feature of precipitation is that its distribution consists
of a discrete component, indicating occurrence of precipitation,
and a continuous one, determining the amount of precipitation. As a
consequence, there are two
basic statistical modeling approaches. The continuous and the discrete
part are either
modeled separately [\citet{CoSt82}, \citet{Wi99}] or together
[\citet{Be87}, \citet{Wi90}, \citet{BaPl92},
\citet{Hu95}, \citet{SaGu04}].
Typically, in the second approach, the distribution of the rainfall
amounts and the probability of rainfall are determined together using
what is called a
censored distribution. Originally, this idea
goes back to \citet{To58} who analyzed household expenditure
on durable goods. For modeling precipitation, \citet{St73} took
up this idea
and modified it by including a power-transformation for the nonzero
part so that the model
can account for skewness.

\subsection{Correlations in space and time}
For modeling processes that involve dependence over space and time, there
are two basic approaches [see, e.g., \citet{CrWi11}]: one which
models the space--time covariance structure without distinguishing
between the time and
space dimensions, and a dynamic one which takes the natural ordering in the
time dimension into account.

The first approach usually follows the traditional geostatistical
paradigm of assuming a parametric
covariance function [for an introduction into geostatistics, see, e.g.,
\citet{Cr93} or \citet{GeEtAl10}]. Several parametric
families specifying explicitly the joint space--time covariance
structure have been proposed [\citet{JoZh97}, \citet{CrHu99},
\citet{Gn02}, \citet{Ma03}, \citet{St05},
\citet{PaSc06}]. Interpretability and, especially, computational
complexity are challenges when working with parametric space--time covariance
functions.

There is, however, a fundamental difference
between the spatial and the temporal dimensions. Whereas there
is an order in the time domain, there exists no obvious order for
space. It is therefore natural to assume a dynamic temporal evolution
combined with a spatially correlated error term [\citet{SoSw96},
\citet{WiCr99}, \citet{HuHs04}, \citet{XuWiFo05},
\citet{GeEtAl05}]. As
\citet{WiHo10} state, the dynamic
approach can be used to construct realistic space--time dependency structures
based on physical knowledge. Further, the temporal Markovian structure
offers computational
benefits.

\subsection{Models for precipitation}
Isham and Cox (\citeyear{IsCo94}) state that there are three broad types of mathematical
models of rainfall: deterministic meteorological
models [\citet{Ma86}], intermediate stochastic models
[\citet{Le61}, \citet{CoIs88}, \citet{WaGuRo84}], and
empirical
statistical models. Meteorological models represent as
realistically as possible the physical processes involved. As noted
by \citet{KyJo99}, deterministic models typically require a large number
of input parameters that are difficult to determine, whereas stochastic
models are usually based on a small number of parameters. Nevertheless,
statistical
models can also incorporate knowledge about physical
processes. Parametrizations can be chosen based on physical knowledge and
covariates reflecting information about the physical processes can be
included.

In the following, we briefly review statistical models for
precipitation. For modeling daily precipitation at a single measuring
site, \citet{StCo84} use a
nonstationary second-order Markov chain to describe
precipitation occurrence and a gamma distribution to describe rainfall
amounts. \citet{HuGu94} and \citet{HuGuCh99} model
precipitation occurrence
using a nonhomogeneous hidden Markov model. With the help of an unobserved
weather state they link large scale atmospheric circulation patterns with
the local precipitation process. \citet{BeHuGu00} and \citet
{ChBaHu99}
both extend this approach by
also modeling precipitation amounts. The former propose to use gamma
distributions, whereas the latter use empirical
distribution functions. \citet{AiThTh09} present a hidden Markov model
in combination with the transformed and censored Gaussian distribution
approach used in \citet{BaPl92}. Also building on the same censoring
idea, \citet{SaGu99} model precipitation occurrence and amount of
precipitation using a transformed multivariate Gaussian model with a spatial
correlation structure. Further works on statistical precipitation
modeling include
Sans{\'o} and Guenni (\citeyear{SaGu99.2,SaGu00}), \citet{BrDiLoYo01},
\citet{StBa02}, \citet{AlGl03},
\citet{SlRaGn07}, \citet{BeRaGn08} and \citet{FuReLe08}.

\subsection{Outline}
The model presented in the following is a hierarchical\break
Bayesian model for spatio-temporal data. At the data stage, we opt for
a modeling approach that determines the discrete and the continuous
parts of
the precipitation distribution together. This is done by assuming the
existence of a latent Gaussian variable which can be interpreted as a
precipitation potential. The mean of the Gaussian variable
is related to covariates through a regression term. The advantages of
this one-part modeling strategy are
twofold: the model contains less parameters and it can deal with the
so-called spatial (and temporal) intermittence effect
[\citet{BaPl92}] which suggests smooth transitions between wet
and dry
areas. This means that at the edge of a dry area the amount
of rainfall should be low. \citet{Wi98} notes that, indeed,
lower rainfall intensity is observed when more neighboring stations are
dry. This feature also reflects the idea that
if a model determines a low probability of rainfall for a given situation,
it should also give a small expected value for its amount
conditional on
this event, and vice versa. However, we note that there is no consensus
in the literature whether
the two parts of precipitation should be modeled together or separately.

At the process level, we use a dynamic model for accounting for
spatio-temporal variation. The model explicitly incorporates knowledge
about the underlying physical processes that determine rainfall,
such as advection, diffusion and convection. Approximating an
integrodifference equation, we obtain a
temporally autoregressive convolution with spatially colored and
temporally white
innovations. The model is nonstationary, anisotropic, and it
allows for nonseparable covariance structures, that is, covariance structures
where spatial and temporal variation interact. While our approach
builds on existing models, it includes
several novel features. With
the help of the Voronoi tessellation, a natural parametrization for
data lying on an
irregular grid is obtained. The
parametrization based on this tessellation is
space as well as time resolution consistent, physically
realistic and allows for modeling irregularly spaced data in a natural
way. To our knowledge, the use of the Voronoi tessellation for
spatio-temporal data on an irregular grid is new. By linking the advection
parameter of the kernel to an external wind vector, the model
is temporally nonstationary.

The model is applied to predict three-hourly precipitation. The
prediction model is based on three forecasted meteorological
variables obtained from an NWP model
as well as past rainfall observations. We compare predictions from the
statistical model with the precipitation forecasts obtained from the NWP.

The remainder is organized as follows. In Section~\ref{SecMod} the model
specifications are presented. In Section~\ref{SecFit} it is shown
how the model can be fitted to data using a Markov chain Monte Carlo (MCMC)
algorithm and how predictions can be obtained. Next, in Section
\ref{SecApl} the model is applied to obtain short term predictions of
three-hourly rainfall. Conclusions are given in Section~\ref{SecCon}.

\section{The model}\label{SecMod}
It is assumed that the rainfall $Y_t(\matt{s})$ at time $t$ on
site $\matt{s}=(x,y)'\in\mathbb{R}^{2}$ depends
on a latent normal variable $W_t(\matt{s})$ through
%
%
\begin{eqnarray}
\label{rainrel} Y_t(\matt{s})&=&0\qquad\mbox{if } W_t(
\matt{s}) \leq0
\nonumber\\[-8pt]\\[-8pt]
&=&W_t(\matt{s})^{\lambda}\qquad \mbox{if } W_t(
\matt{s})>0,\nonumber
\end{eqnarray}
where $\lambda>0$. A power transformation is needed since precipitation
amounts are
more skewed than a truncated normal distribution and since
the scatter of the precipitation amounts increases with the
average amount. The latent variable $W_t(\matt{s})$ can be interpreted
as a precipitation
potential.

This latent variable $W_t(\matt{s})$ is modeled as a Gaussian process that
is specified as
%
%
\begin{equation}
\label{model1} W_t(\matt{s})=\matt{x}_t(
\matt{s})^T\vect{\beta}+\xi_t(\matt{s})+
\nu_t(\matt{s}),
\end{equation}
where $\bolds{\beta} \in\mathbb{R}^{k}$,
and $\nu_t(\matt{s})\sim
N(0,\tau^2), \tau^2>0$, are i.i.d. The mean
$\matt{x}_t(\matt{s})^T\vect{\beta}$ of $W_t(\matt{s})$ is assumed
to depend
linearly on regressors $\matt{x}_t(\matt{s})\in\mathbb{R}^{k}$. For
notational convenience, we split the terms specifying the covariance
function into a structured
part $\xi_t(\matt{s})$ and an unstructured ``nugget'' $\nu_t(\matt
{s})$. The term
$\xi_t(\matt{s})$ is a zero-mean Gaussian process that accounts for
structured variation in time and space. It is specified below in Section
\ref{TSDep}. The nugget $\nu_t(\matt{s})$ models microscale variability
and measurement
errors. Since, typically, the resolution of the data does not allow for
distinguishing between microscale variability and measurement
errors, we model these two sources of variation together. Note that the
covariates $\matt{x}_t(\matt{s})$ will usually be time and location
dependent. In addition to weather characteristics, Fourier harmonics
can be included to account for
seasonality, and functions of coordinates can account for smooth effects
in space.

\subsection{The convolution autoregressive model}\label{TSDep}

We follow the dynamic approach and define an explicit time evolution
through the following integrodifference equation (IDE):
%
%
\begin{equation}
\label{IDE} \xi_t(\matt{s})=\int_{\mathbb{R}^{2}}{h_{\vect{\vartheta}}
\bigl(\matt{s} -\matt{s}'\bigr)\xi_{t-1}\bigl(
\matt{s}'\bigr)\,d\matt{s}'}+\varepsilon_{t}(
\matt{s}),\qquad \matt{s}\in\mathbb{R}^{2},
\end{equation}
where $\varepsilon_{t}(\matt{s})$ is a Gaussian innovation
that is white in time and colored in space, and $h_{\vect{\vartheta}}$
is a
Gaussian kernel,
%
%
\begin{equation}
\label{kernel} h_{\vect{\vartheta}}\bigl(\matt{s}-\matt{s}'\bigr)= \phi
\exp\bigl(-\bigl(\matt{s}-\matt{s}'- \vect{\mu}_t
\bigr)^T \mat{\Sigma}^{-1}\bigl(\matt{s}-
\matt{s}'- \vect{\mu}_t\bigr) \bigr),
\end{equation}
where the parameter vector ${\vect{\vartheta}}$ combines $\phi$ and the
elements of $\vect{\mu}_t$ and $\mat{\Sigma}^{-1}$. Note that
$\vect{\mu}_t$
shifts the kernel and $\mat{\Sigma}^{-1}$ determines the
range and the degree of anisotropy. The parameter $\phi$ controls the
amount of temporal correlation. More details on the interpretation of the
model and specific choices of the parameters $\vect{\mu}_t$ and
$\mat{\Sigma}$ are discussed below in Section~\ref{Parametrization}. An
illustration of this kernel can be found in the application in Section
\ref{FitRes}.

In the following, we assume that we
have $N$ measurement locations
$\matt{s}_i$, $i=1,\ldots,N$, where measurements are made at times
$t=1,\ldots,T$. Instead of working with a fine
spatial grid with many missing observations, we formulate an approximate\vadjust{\goodbreak}
model for the values at the stations only,
$\vect{\xi}_t=(\xi_t(\matt{s}_1),\ldots,\xi_t(\matt{s}_N))'$. Discretizing
the integral in (\ref{IDE}), we obtain
%
%
\begin{eqnarray}
\label{CKTARApprox} \int_{\mathbb{R}^2}h_{\vect{\vartheta}}\bigl(
\matt{s}_i-\matt{s}'\bigr)\xi_{t-1}\bigl(
\matt{s}'\bigr) \,d\matt{s}' &\approx& \int
_{A}h_{\vect{\vartheta}}\bigl(\matt{s}_i-
\matt{s}'\bigr)\xi_{t-1}\bigl(\matt{s}'\bigr)\,d
\matt{s}'
\nonumber\\[-8pt]\\[-8pt]
&\approx& \sum_{j=1}^N{h_{\vect{\vartheta}}(
\matt{s}_i-\matt{s}_j)\xi_{t-1}(\matt
{s}_j)|A_j|}.\nonumber
\end{eqnarray}
Here $A \subset\mathbb{R}^2$ is an area which contains the convex
hull of
all stations, the sets $A_i, i=1,\ldots,N$, form a tessellation of $A$ with
$\matt{s}_i \in A_i$ and $|A_j|$ denotes the area of cell $A_j$.

Our model can then be written as the vector autoregression
%
%
\begin{equation}
\label{model2} \vect{\xi}_t=\phi\matt{G}_t \vect{
\xi}_{t-1}+\vect{\varepsilon}_t,\qquad \matt{G}_t
\in\mathbb{R}^{N \times N},
\end{equation}
where
%
%
\begin{equation}
\label{kernel2} (\matt{G}_t )_{ij}=\exp\bigl(-\bigl(
\matt{s_i}-\matt{s}_j'- \vect{
\mu}_t\bigr)^T \mat{\Sigma}^{-1}\bigl(
\matt{s}_i-\matt{s}_j'- \vect{
\mu}_t\bigr) \bigr)\cdot|A_j|,
\end{equation}
and where $\vect{\varepsilon}_t=(\varepsilon_t(\matt{s}_1),\ldots,
\varepsilon_t(\matt{s}_N))'$.

Note that this process does not exhibit explosive growth if the largest
eigenvalue of $\phi\matt{G}_t$ is smaller than one. To ensure this,
we check
in our application that the largest eigenvalue is smaller than one for the
parameters at the posterior modes.

%
%
\begin{figure}

\includegraphics{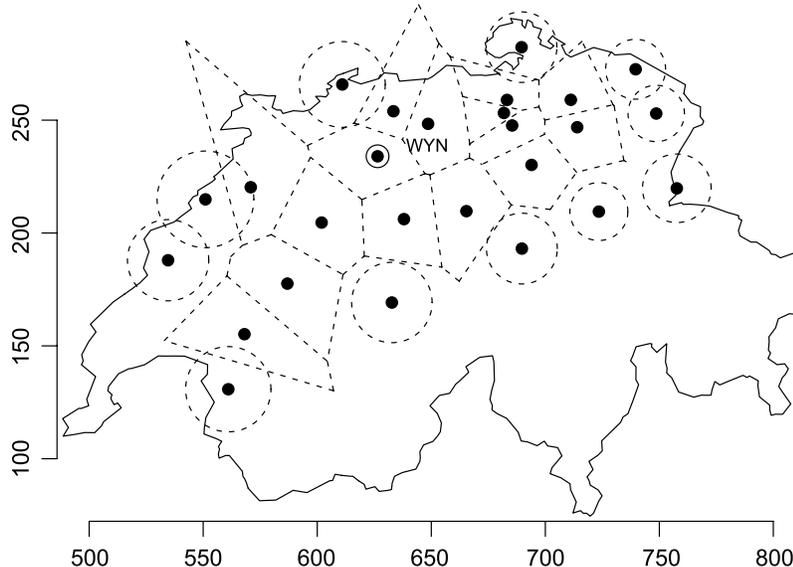}

\caption{Locations of stations. Both axes are in km using the Swiss
coordinate system (CH1903). The lines
illustrate the Voronoi tessellation. Cells with unbounded area have been
replaced by circles whose area is determined as described in the text.}
\label{fig:Stations}
\end{figure}

If the $\matt{s}_i$'s form a regular grid, a tessellation is straightforward.
Otherwise, we propose to use the Voronoi tessellation
[\citet{Vo08}] which decomposes the space. Specifically, each site
$\matt{s}_i$ has a corresponding Voronoi cell consisting of all points
closer to $\matt{s}_i$ than to
any other site $\matt{s}_j$, $j \neq i$ [see, e.g., \citet
{OkBo00} for more
details]. Stations on the boundary of the convex hull have cells with
infinite area. For these stations, we define $|{A}_i|$ as described
in the following. We first calculate the Voronoi
tessellation of $\mathbb{R}^2$. We then replace unbounded cells by
cells whose area is the average area of the neighboring bounded cells.
In Figure~\ref{fig:Stations}, the Voronoi tessellation for the Swiss
stations used in the application below is shown as an example.
Concerning
the stations on the boundary, 
the circles
represent the surface area $|{A}_i|$.

As mentioned before, the $\vect{\varepsilon}_t$'s are assumed to be
independent
over time and colored in space. More precisely, we assume a stationary,
isotropic Gaussian random field
%
%
\begin{equation}
\vect{\varepsilon}_t\sim N\bigl(0,\sigma^2
\matt{V}_{\rho_0}\bigr), \qquad\sigma^2>0,
\end{equation}
with
%
%
\begin{equation}
\label{CovDef} { (\matt{V}_{\rho_0} )}_{ij}=\exp
(-d_{ij}/\rho_0 ),\qquad \rho_0>0,\qquad 1 \leq i,j, \le
N,
\end{equation}
where $d_{ij}$ denotes the Euclidean distance between two sites $i$ and
$j$. The
exponential correlation function is used for computational convenience. In
principle, it is possible to use other covariance functions, for
instance, other members of the Mat\'{e}rn family.

The approximation in (\ref{CKTARApprox}) assumes that
$h_{\vect{\vartheta}}$ is approximately constant in each cell. If
some cells
are considered to be too large for this approximation to be reasonable,
additional points $\matt{s}_j^\ast$ can be added for which all observations
are missing. Since such additional points increase the computational load,
some compromise has to be found between accuracy and computational
feasibility.

\subsection{Interpretation and parametrization of the kernel
function}\label{Parametrization}
For the purpose of interpretation, we note that, in the limit when the
temporal spacing goes to zero,
the solution of the IDE (\ref{IDE}) can also be
written as the solution of the stochastic partial differential equation
(SPDE) [see \citet{BrKa00}]
%
%
\begin{equation}
\label{SPDE1} \frac{\partial}{\partial
t}\xi_t(\matt{s})=-\vect{
\mu}_t\cdot\nabla\xi_t(\matt{s}) +\frac{1}{4}
\nabla\cdot\mat{\Sigma}\nabla\xi_t(\matt{s}) -\eta
\xi_t(\matt{s})+B_t(\matt{s}),
\end{equation}
where $\nabla= (\frac{\partial}{\partial x},
\frac{\partial}{\partial y} )$ is the gradient operator and where
$B_t(\matt{s})$ is temporally independent and spatially dependent. The
terms have the following interpretations: $\vect{\mu}_t\cdot\nabla
\xi_t(\matt{s})$ models
advection, $\vect{\mu}_t$ being a drift or velocity vector. The second
term is a
diffusion term that can incorporate anisotropy, and $-\eta\xi_t(\matt{s})$
accounts for damping. The damping parameter $\eta$ is related to $\phi$
and $\vect{\Sigma}$ through $\eta=-\log(\phi\pi|\vect{\Sigma
}|^{1/2})$. $B_t(\matt{s})$ is a source-sink or stochastic forcing
term that can be interpreted as modeling convective phenomena. This
interpretation is based on the reasoning that typically convective precipitation
cells emerge and cease on the domain of interest in contrast to larger
scale advective precipitation that is being transported over the area.

We now turn to the discussion of the parameterization of $\vect{\mu
}_t$ and
$\mat{\Sigma}$. In our application, we have information about
wind. It is assumed that
the drift term $\vect{\mu}_t$ is proportional to this external wind
vector. With $\vect{\mu}_t$ varying over time, the model is temporally
nonstationary. It is also conceivable that in certain situations
$\mat{\Sigma}$ or $\eta$ may vary over time and/or space, thus obtaining
different forms of nonstationarity. Concerning $\mat{\Sigma}$, it is
thought that potential anisotropy is related to topography. Denoting by
$\matt{w}_t$ the wind vector at time $t$, we assume
%
%
\begin{eqnarray}
\label{NonIsoDriftConvAR}
\vect{\mu}_t&=&u\cdot\matt{w}_t\quad
\mbox{and}\nonumber\\[-8pt]\\[-8pt]
\mat{\Sigma}^{-1}&=&\frac{1}{\rho_1^2}\pmatrix{\cos{\alpha} &
\sin{\alpha}
\cr
-c\cdot\sin{\alpha} & c\cdot\cos{\alpha}}^{T} \pmatrix{
\cos{\alpha} & \sin{\alpha}
\cr
-c\cdot\sin{\alpha} & c\cdot\cos{\alpha}},\nonumber
\end{eqnarray}
where $u \in\mathbb{R}$, $c > 0$, and $\alpha\in[0,\pi/2]$. We use
a wind vector which is averaged over the entire area, but the wind could
also change locally.
The motivation for writing $\mat{\Sigma}$ in the given form comes from
considering a coordinate transformation
%
%
\begin{equation}
\pmatrix{x'
\cr
y'}= \pmatrix{\cos{\alpha} & \sin{
\alpha}
\cr
-c\cdot\sin{\alpha} & c\cdot\cos{\alpha}} \pmatrix{x
\cr
y},
\end{equation}
where the parameter $\alpha$ is the angle of
rotation, and $c$ determines the degree of anisotropy, $c=1$ corresponding
to the isotropic case.
$\rho_1$ is a range
parameter that determines the degree of interaction between spatial and
temporal correlation. See Section~\ref{FitRes} for an illustration of a
kernel with the
above parametrization.

The resulting model is nonstationary and
incorporates anisotropy. Finally, we note that there are various other
possible choices of parametrizations. For
instance, a relatively simple model can be obtained by assuming
%
%
\begin{equation}
\label{ConvAR} \vect{\mu}_t=\matt{0} \quad\mbox{and}\quad \mat{
\Sigma}^{-1}=\frac{1}{\rho_1^2} \pmatrix{1 & 0
\cr
0 & 1 },
\end{equation}
that is, no drift and an isotropic diffusion term. There is still
spatio-temporal interaction, though, which implies that
the model is not separable in the sense that (\ref{SepCov}) does not hold.
We can simplify further and take not only $\vect{\mu}_t=\matt{0}$,
but also
$\mat{\Sigma}=\matt{0}$, leading to $\matt{G}_t$ being the
identity matrix
%
%
\begin{equation}
\label{STAR} \vect{\xi}_t=\phi\vect{\xi}_{t-1}+\vect{
\varepsilon}_t.
\end{equation}
This means that each point at time $t-1$ only has an
influence on itself at time $t$, that is, there is no spatio-temporal
interaction and
the model is separable.

\subsection{Discussion of the model}

\mbox{}

\textit{Propagator matrix $\matt{G}_t$.}\quad
Using a parametrized propagator matrix $\matt{G}_t$ in (\ref{model2})
has the
obvious
advantage that less parameters are needed than in the general
case, in which each entry in the matrix has to be estimated, resulting
in $N^2$ parameters. Moreover, in contrast to the general case, the
parametric approach allows for making predictions at sites where no
measurements are
available, which
is often of interest in applications.\vspace*{8pt}

\textit{Space resolution consistency.}\quad
At first sight, it might be tempting to use a simpler parametrization
of $\matt G_t$ not based on a convolution but of the form
%
%
\begin{equation}
\label{ARDef} (\matt{G}_t)_{ij}= \exp\bigl(-
(d_{ij}/\rho_1 )^2 \bigr).
\end{equation}
However, such a model has the following important
drawback. Assume, for instance, that a station
$i$ is surrounded by two neighboring sites $j$ and $k$. Say that both
stations $j$ and $k$ lie at the same distance from $i$ but in different
directions. Consequently, $j$ and $k$ at time $t-1$ exercise the same
influence on $i$ at time $t$. If one adds an additional station $l$ very
close to $k$, the joint
influence of $k$ and $l$ at time $t-1$ on site $i$ at time $t$ would then
approximately be twice as big as the one of site $j$. This means that the
distribution of the process at point $i$ depends on the number and the
location of stations in the neighborhood at which it has been observed.
The convolution model, on the
other hand, does not
exhibit this drawback. Furthermore, the convolution model has the advantage
that it is ``space resolution consistent,'' that is, it retains
approximately
its temporal Markovian structure if one, or several, sites are removed
from the
domain. This does not hold true for the simpler vector autoregressive
model as specified in (\ref{ARDef}).\vspace*{8pt}

\textit{Space--time covariance structure.}\quad
In the following, let us turn to the spatio-temporal dependence structure
of the latent process $\vect{\xi}_t$. A random field $\xi_{t}(\matt
{s})$, $(\matt{s},t)\in
\mathbb{R}^{2} \times\mathbb{R}$ is said to have a \textit{separable}
covariance structure [\citet{GnGeGu07}] if there
exist purely spatial and purely temporal covariance functions $C_S$ and
$C_T$, respectively, such that
%
%
\begin{equation}
\label{SepCov} \mbox{cov}\bigl(\xi_{t_1}(\matt{s}_1),
\xi_{t_2}(\matt{s}_2)\bigr)=C_S(\matt
{s}_1,\matt{s}_2)\cdot C_T(t_1,t_2).
\end{equation}
The convolution based approach allows for nonseparable covariance structures,
whereas the separable autoregressive model in (\ref{STAR}) has a
separable covariance
structure.\vspace*{8pt}

\textit{Extremal events.}\quad For the data model as specified in
equation (\ref{rainrel}),
Hern{\'a}n\-dez, Guenni and Sans{\'o} (\citeyear{HeGuSa09}) showed that the
distribution of the maxima is a Gumbel. If the focus lies on extremal
events, other distributions,
which have Fr\'echet maxima, can be used, for instance, a $t$-distribution.
The $t$-distribution is particularly attractive since
it is a scale mixture of
normal distributions.
To be more specific, if $S_t$ has a $\chi_{df}^2$
distribution, then
$\matt{W}_t=\matt{x}_t^T\vect{\beta}+
(\vect{\xi}_t+\vect{\nu}_t)/\sqrt{S_t/df}
$ has a multivariate $t$-distribution.\vadjust{\goodbreak} This means that the fitting
algorithm introduced below can be extended to the $t$-distribution case
by introducing an additional latent variable $S_t$.

\section{Fitting and prediction}\label{SecFit}
Fitting is done using a Markov chain Monte Carlo method (MCMC), the
Metropolis--Hastings algorithm [\citet{MeetAl53},
\citet{Ha70}]. Concerning most parameters, it will be shown that the
full conditionals are known distributions. Therefore, Gibbs sampling
[\citet{GeSm90}] can be used in these cases.

For convenience and later use, we combine the parameters characterizing the
model into a vector
$\vect{\theta}=(\lambda, \vect{\beta}', \tau^2, \sigma^2, \rho_0, \vect
{\vartheta}')'$
and call them \textit{primary parameters.} Our goal is to simulate from
the joint posterior distribution of
these parameters and the latent variables
$\vect{\xi}=(\vect{\xi}_1,\ldots,\vect{\xi}_T),\vect{\xi}_0$,
and
$\matt{W}=(\matt{W}_1,\ldots,\matt{W}_T)$. We note that those
$W_t(\matt{s}_i)$ that correspond to observed values above zero are known.
In that case the full conditional distribution
consists of a Dirac distribution at $Y_t(\matt{s}_i)^{1/\lambda}$. For
handling the censored values and for allowing for missing values, we
adopt a
data augmentation approach [\citet{SmRo93}] as specified below in
equation (\ref{DataAugm}). See Section~\ref{FullCond} for more details.

Assuming prior independence among the
primary parameters,
the prior distributions are specified as
%
%
\begin{eqnarray}
&&
P\bigl[\lambda,\bolds{\beta}, \tau^2, \sigma^2,
\rho_0, \phi, u, \rho_1, \alpha, c, \vect{
\xi}_0\bigr] \nonumber\\[-8pt]\\[-8pt]
&&\qquad\propto\frac{1}{\tau^2}\frac{1}{\sigma^2}P[
\rho_0]P[\rho_1] P[u] P[c]P[\alpha] P \bigl[\vect{
\xi}_0|\sigma^2,\rho_0 \bigr]\nonumber
\end{eqnarray}
with $\vect{\xi}_0$ having a normal
prior $P[\vect{\xi}_0|\sigma^2,\rho_0]=
N(0,\sigma^2 \matt{V}_{\rho_0})
$.
Further, $\rho_0$ and $\rho_1$ have gamma priors with mean
$\mu_{\rho}$ and variance $\sigma^2_{\rho}$. For $c$, we assume a gamma
prior with mean $1$ and
variance $1$, $\alpha$ has a uniform prior on $[0,\pi/2]$, and $u$
has a
normal prior with mean $0$ and variance $10^4$.
Further, we assume locally uniform priors on
$\log(\tau^2)$ and $\log(\sigma^2)$ as well as for
$\phi$, $\lambda$ and $\vect{\beta}$.

In our application, we choose to use informative priors
for $\rho_0$ and $\rho_1$. It is known that in model-based
geostatistics difficulties can arise when estimating the variance and
scale parameters of the
exponential covariogram [see, e.g., \citet{WaRi87},
\citet{MaWa89}, \citet{DiTaMo98}]. For the geostatistical
covariance model, \citet{Zh04} shows that
the product of the two parameters can be estimated consistently, and
\citet{St90} shows that it is the product
of the two parameters that matters more than the
individual parameters for spatial interpolation. Further, \citet
{BeEtAl01} show that, at least in
the simplest setting, the posterior of the range parameters is improper for
most noninformative priors. Given these considerations, we think that
using informative priors for the two range parameters $\rho_0$ and
$\rho_1$
is appropriate. In our example, we chose priors with
mean $\mu_{\rho}=100$ and variance $\sigma^2_{\rho}=10$. We have tried\vadjust{\goodbreak}
different informative priors. The less informative they are, the worse
are the mixing properties of the MCMC algorithm. In line with the
results of \citet{St90} and \citet{Zh04}, we have made the
experience
that different choices of priors on these range parameters do not have
a strong impact on the predictive performance of the model.

The posterior distribution is then proportional to
%
%
\begin{eqnarray}
\label{fullcondrain}
&& \biggl(\frac{1}{\sigma^2} \biggr)^{
{N(T+1)}/{2}+1} \biggl(
\frac
{1}{\tau^2} \biggr)^{{NT}/{2}+1}|\matt{V}_{\rho_0}|^{-
({T+1})/{2}}
\prod_{Y_t(\matt{s}_i)>0}\frac{Y_t(\matt{s}_i)^{1/\lambda-1}}{\lambda
}
\nonumber
\\
&&\qquad{}\times\exp\Biggl(-\frac{1}{2} \sum_{t=1}^T
\frac{1}{\tau^2}\bigl\|\matt{W}_t-\matt{x}_t^T
\vect{\beta}-\vect{\xi}_t\bigr\|^2
\nonumber\\[-8pt]\\[-8pt]
&&\qquad\hspace*{34.5pt}{}+\frac{1}{\sigma^2} (\vect{\xi}_t-\phi\matt{G}_t
\vect{\xi}_{t-1} )'\matt{V}_{\rho_0}^{-1} (
\vect{\xi}_t-\phi\matt{G}_t \vect{\xi}_{t-1} )
\Biggr)
\nonumber
\\
&&\qquad{}\times\exp\biggl(-\frac{1}{2}\frac{1}{\sigma^2}\vect{\xi
}_0'\matt{V}_{\rho_0}^{-1}\vect{
\xi}_0 \biggr)\cdot P[\rho_0] \cdot P[\vect{\vartheta}]
\cdot\mathbf{1}_{\{
W_t(\matt{s}_i)\leq0\ \forall i,t\dvtx Y_t(\matt{s}_i)=0\}}.\nonumber
\end{eqnarray}
The product in the first line is
the Jacobian for the power transformation in
(\ref{rainrel}).
Note that missing observations do not cause any problem.
If $Y_t(\matt{s}_i)$ is missing, there is no respective term in the product
nor a corresponding condition for the indicator function.

\subsection{Full conditional distributions}\label{FullCond}
In the following, we derive full conditional
distributions for the individual parameters.

It is readily seen that the full conditional of $\vect{\beta}$ is a
multivariate normal distribution, and the full conditional distribution of
$\phi$ is a normal distribution as well.
The full conditionals of both $\sigma^2$ and $\tau^2$ are inverse gamma
distributions.

For obtaining the full conditionals of $\matt W_t$,
we partition
its components according to whether
$Y_t(\matt{s}_i)$ is above zero, equal to zero, or missing.
Denote by
$i_t^{[+]}$ those
indices for which $Y_t(\matt{s}_i)>0$, by $i_t^{[0]}$
those with $Y_t(\matt{s}_i)=0$, and by $i_t^{[m]}$ the missing ones.
The vector $\matt{W}_t$ can then be
partitioned into $\matt{W}_t^{[+]}$, $\matt{W}_t^{[0]}$, and
$\matt{W}_t^{[m]}$ accordingly.
We remark that $\matt{W}_t^{[0]}$ and $\matt{W}_t^{[m]}$ are latent
variables, whereas $\matt{W}_t^{[+]}$ corresponds to transformed observed
values. In addition, $\matt{W}_t^{[0]}$ has the restriction that all its
values must be smaller than zero, $\matt{W}_t^{[0]}\leq\matt{0}$. For
facilitating understanding, we note that $W_t(\matt{s}_i)$ can be
written as
%
%
\begin{eqnarray}
\label{DataAugm} W_t(\matt{s}_i)&=&W_{t}^{[+]}(
\matt{s}_i)=Y_t(\matt{s}_i)^{1/\lambda}
\qquad\mbox{if } Y_t(\matt{s}_i)>0
\nonumber
\\
&=&W_{t}^{[0]}(\matt{s}_i)\qquad \mbox{if }
Y_t(\matt{s}_i)=0
\\
&=&W_{t}^{[m]}(\matt{s}_i)\qquad\mbox{if }
Y_t(\matt{s}_i) \mbox{ is missing}.
\nonumber
\end{eqnarray}
The full conditional of $\matt{W}_t^{[m]}$ is
then a multivariate normal distribution with mean and
covariance
%
%
\begin{equation}
\mu_{\matt{W}_t^{[m]}}=\bigl(\matt{x}_t^T\vect{\beta}+
\vect{\xi}_t\bigr)^{[m]} \quad\mbox{and}\quad\mat{
\Sigma}_{\matt{W}_t^{[m]}}=\tau^2\cdot\matt{I}.
\end{equation}
Similarly, the full conditional
distribution of $\matt{W}_t^{[0]}$ is a truncated multivariate normal
distribution with mean and
covariance
%
%
\begin{equation}
\mu_{\matt{W}_t^{[0]}}=\bigl(\matt{x}_t^T\vect{\beta}+
\vect{\xi}_t\bigr)^{[0]}\quad\mbox{and}\quad \mat{
\Sigma}_{\matt{W}_t^{[0]}}=\tau^2\cdot\matt{I}.
\end{equation}
As mentioned before, the full conditional of
$\matt{W}_t^{[+]}$ is a Dirac distribution with point mass at $
(\matt{Y}_t^{[+]} )^{1/\lambda}$.

Concerning the latent variables $(\vect{\xi}_{0},\vect{\xi}_1
,\ldots,
\vect{\xi}_T)$, we note
that conditional on~$\vect{\theta}$, $(\vect{\xi}_t,\matt{W}_t)$
is a
linear Gaussian state space model. Therefore, a
sample from the joint full conditional of
$(\vect{\xi}_{0},\vect{\xi}_1,\ldots,\vect{\xi}_T)$ can be
obtained using the forward filtering backward sampling (FFBS) algorithm proposed
by \citet{CaKo94} and \citet{Fr94}. The forward filtering step
corresponds to the Kalman filter [see, e.g., \citet{WeHa97} and
\citet{Ku01}].

Alternatively, one can also use single $t$ updates. The full
conditional of one~$\vect{\xi}_t$, $0\leq t \leq T$, is a normal
distribution
$N (\vect{\mu}_{\vect{\xi}_t},\mat{\Sigma}_{\vect{\xi
}_t} )$. In
the case of the separable model, the mean $\vect{\mu}_{\vect{\xi}_t}$
depends on
$\vect{\xi}_{t-1}$ and $\vect{\xi}_{t+1}$, whereas the covariance
matrix $\mat{\Sigma}_{\vect{\xi}_t}$
does not depend
on $t$. This is convenient for simulation since
its Cholesky decomposition
has to be calculated only once in each update cycle.
In contrast, in the sampling step of the FFBS
algorithm, one has to calculate
a Cholesky decomposition
for each $t$. The advantage that the FFBS algorithm mixes
better than the single $t$ update algorithm per update cycle is outweighed
by the fact that an update cycle of
the single $t$ update algorithm is a lot faster than one of the FFBS
algorithm. Thus, more
effective samples can be obtained with the single $t$ update algorithm
per time. In the case of the nonstationary
anisotropic drift model, though, $\mat{\Sigma}_{\vect{\xi}_t}$ in the
single $t$ update algorithm is not
constant over time. Thus, a Cholesky decomposition needs to be computed for
each $t$ anyway, meaning that the FFBS algorithm is preferable.

In summary, we made the experience that it is recommendable to use single
$t$ updates for temporally stationary models where the covariance
$\mat{\Sigma}_{\vect{\xi}_t}$ of the full conditional of one $\vect
{\xi
}_t$ is constant over time. If
$\mat{\Sigma}_{\vect{\xi}_t}$ changes over time, we recommend using the
FFBS algorithm.

For the remaining parameters, that is, $\rho_0$, $\vect{\vartheta}$
(excluding $\phi$)
and $\lambda$, there
is no apparent distribution family from which one can simulate.
Therefore, Metropolis
steps will be used.
We note that the full conditional distribution of $\lambda$ is proportional
to
%
%
\begin{equation}\quad
\prod_{Y_t(\matt{s}_i)>0} \biggl(\frac{Y_t(\matt{s}_i)^{1/\lambda
-1}}{\lambda} \biggr)\exp
\biggl(-\frac{1}{2} \sum_{Y_t(\matt{s}_i)>0}
\frac{1}{\tau^2}\bigl\|Y_t(\matt{s}_i)^{1/\lambda}-
\matt{x}_t^T\vect{\beta}-\vect{\xi}_t\bigr\|^2
\biggr).
\end{equation}
The parameter $\lambda$ is sampled on the log-scale. This
means that we first transform it to the log scale. Then a
proposal is obtained by sampling from a normal distribution with the mean
equal to the last value of the parameter. Thereafter, this
proposal is accepted with a probability that is given by the usual
Metropolis--Hasting algorithm [see, e.g., \citet{ChGr95}].

Finally, $\rho_0$ and $\vect{\vartheta}$ (excluding $\phi$) are sampled
together.
The full conditional is proportional to
%
%
\begin{eqnarray}
&&\exp\Biggl(-\frac{1}{2\sigma^2} \Biggl( \sum_{t=1}^T
(\vect{\xi}_t-\phi\matt{G}_t \vect{\xi}_{t-1}
)'\matt{V}_{\rho_0}^{-1} (\vect{\xi}_t-
\phi\matt{G}_t \vect{\xi}_{t-1} )+\vect{
\xi}_0'\matt{V}_{\rho
_0}^{-1}\vect{
\xi}_0 \Biggr) \Biggr)\nonumber\\[-8pt]\\[-8pt]
&&\qquad{}\times|\matt{V}_{\rho_0}|^{-({T+1})/{2}}.\nonumber
\end{eqnarray}

\subsection{Prediction}\label{Prediction}
We consider predictions at new locations and/or times as well as
predictions of areal averages. It turns out that in the case of areal
averages, the Voronoi tessellation is again useful.

One way to obtain predictions is to augment
the data $\matt{Y}_{\mathrm{obs}}$ with missing values at the locations or times
where predictions are made.
When doing so,
the MCMC algorithm implicitly draws from the corresponding predictive
distribution. See the previous Section~\ref{FullCond} on how to handle
missing values.

If one does not specify the points in space and time where predictions are
to be made prior to model fitting, the predictive distribution of a new
set of observations
$\matt{Y}^*=(Y^*_{t^*_1}(\matt{s}^*_1),\ldots,Y^*_{t^*_k}(\matt
{s}^*_k))'$ is
calculated as
%
%
\begin{eqnarray}
P\bigl[\matt{Y}^*|\matt{Y}_{\mathrm{obs}}\bigr]&=&\int P\bigl[\matt{Y}^*|\vect{
\xi}^*,\vect{\theta} \bigr]P\bigl[\vect{\xi}^*|\vect{\xi},\vect{\theta
}\bigr] P[
\vect{\xi},\vect{\theta}|\matt{Y}_{\mathrm{obs}}]\,d\vect{\xi}^*\,d\vect{\xi}\,
d\vect{
\theta}
\nonumber
\\
&\approx&\frac{1}{m} \sum_{i=1}^m
\int P\bigl[\matt{Y}^*|\vect{\xi}^*,\vect{\theta}^{(i)}\bigr]P\bigl[
\vect{\xi}^*|\vect{\xi}^{(i)},\vect{\theta}^{(i)}\bigr]\,d\vect{
\xi}^*
\\
&\approx&\frac{1}{m} \sum_{i=1}^m
P\bigl[\matt{Y}^*|\vect{\xi}^{*(i)},\vect{\theta}^{(i)}\bigr],\nonumber
\end{eqnarray}
where $\matt{Y}_{\mathrm{obs}}$ denotes the observed data, $\vect{\xi}$ and
$\vect{\xi}^*$ the
latent Gaussian process at the observed and predicted sites,
respectively, and $\vect{\theta}$ all the remaining
parameters. Samples $\vect{\theta}^{(i)}$ and
$\vect{\xi}^{(i)}, i=1,\ldots,m$, from their posterior distribution are
obtained by the MCMC algorithm, and $\vect{\xi}^{*(i)}$
is sampled from $P[\vect{\xi}^*|\vect{\xi}^{(i)},\vect{\theta}^{(i)}]$.

When $\vect{\xi}^*$ is modeled at the same sites as $\vect{\xi}$
but at
different time points, the distribution
$P[\vect{\xi}^*|\vect{\xi}^{(i)},\vect{\theta}^{(i)}]$ is
Gaussian and readily obtained using (\ref{model2}).

In the case when predictions are made at unobserved
sites $s \in S$ and time $t$,
$P[\vect{\xi}_t^*|\vect{\xi},\vect{\theta}]$ can be calculated as
described in the following. First, because of the temporal Markov
property, $P[\vect{\xi}_t^*|\vect{\xi},\vect{\theta}]$ is equal to
$P[\vect{\xi}_t^*|\vect{\xi}_{t-1},\vect{\xi}_t,
\vect{\xi}_{t+1},\vect{\theta}]$. This density is then obtained by
considering the augmented model
%
%
\begin{eqnarray}
\label{ExtMod}
\pmatrix{\vect{\xi}_t
\cr
\vect{\xi}_t^*}&=&
\phi\pmatrix{\matt G_t
\cr
\matt G^*_t}\vect{
\xi}_{t-1}+\pmatrix{\vect{\varepsilon}_t
\cr
\vect{
\varepsilon}_t^*},\nonumber\\[-8pt]\\[-8pt]
\vect{\xi}_{t+1}&=&\phi\bigl(\matrix{\matt
H_{t+1} & \matt H_{t+1}^*}\bigr)\pmatrix{\vect{\xi}_t
\cr
\vect{\xi}_t^*}+\vect{\varepsilon}_{t+1},\nonumber
\end{eqnarray}
where $\matt G^*_t$ is defined analogously to (\ref{kernel2}), $\matt
H_{t+1}$ and $\matt H_{t+1}^*$ are obtained from the same
approximations as
in (\ref{CKTARApprox}), and the covariances of $\vect{\varepsilon
}_t$ and
$\vect{\varepsilon}_t^*$ are as in (\ref{CovDef}). By (\ref
{ExtMod}), the
conditional distribution of
$\vect{\xi}_t,\vect{\xi}_t^*,\vect{\xi}_{t+1}$ given
$\vect{\xi}_{t-1}$ is normal. Therefore, also the conditional
distribution of $\vect{\xi}_t^*$ given
$\vect{\xi}_{t-1},\vect{\xi}_t,\vect{\xi}_{t+1}$ is
Gaussian. Its mean and covariance can be computed by noting that
%
%
\begin{eqnarray}
\label{FirstEq} P\bigl[\vect{\xi}_t^*|\vect{\xi}_{t-1},
\vect{\xi}_t,\vect{\xi}_{t+1},\vect{\theta}\bigr]&\propto&
P\bigl[\vect{\xi}_{t+1}|\vect{\xi}_t^*,\vect{
\xi}_t,\vect{\theta}\bigr]P\bigl[\vect{\xi}_t^*|\vect{
\xi}_{t-1},\vect{\xi}_t,\vect{\theta}\bigr]
\nonumber\\[-8pt]\\[-8pt]
&\propto&P\bigl[\vect{\xi}_{t+1}|\vect{\xi}_t^*,\vect{
\xi}_t,\vect{\theta} \bigr]P\bigl[\vect{\xi}_t,\vect{
\xi}_t^*|\vect{\xi}_{t-1},\vect{\theta}\bigr]\nonumber
\end{eqnarray}
and then completing the square in the exponent of the last expression.

In many cases, for instance, when the focus lies on flooding, areal
averages
%
%
\begin{equation}
\label{AaerAvg} \bar{Y}^{(A^*)}_t=\frac{1}{|A^*|}\int
_{A^*} Y_t(\matt{s}) \,d\matt{s}
\end{equation}
of precipitation are of interest. If $Y_t(\matt{s})$ is observed on an
irregular grid, one could first define
a regular grid, then interpolate the nonobserved grid points, and
approximate the integral in (\ref{AaerAvg}) by a Riemann sum. However,
since the regular
grid usually becomes very large, this is computationally
expensive. Instead, we propose to use the
Voronoi tessellation once again to approximate the integral
%
%
\begin{equation}
\label{AaerAvg2} \bar{Y}^{(A^*)}_t=\frac{1}{|A^*|}\int
_{A^*} Y_t(\matt{s}) \,d\matt{s}\approx
\frac{1}{|A^*|}\sum_{j=1}^NY_t(
\matt{s}_j)\bigl|A_j\cap A^*\bigr|.
\end{equation}
Thereby, an adequate weight
$|A_j\cap A^*|$
is given to each station.
Samples
from the predictive distribution of $\bar{Y}^{(A^*)}_t$ can be obtained
by simulating
$Y^{(i)}_t(\matt{s}_j)$ from their predictive distribution and inserting
them in (\ref{AaerAvg2}).

We note that the areal prediction becomes
deterministic
if all $Y_t(\matt{s}_j)$ consist of observed values. This
means that uncertainty about values of $Y_t(\matt{s})$ at locations
where no observations are made is
implicitly ignored with the above approximation. This can be
amended for by first making predictions at a few sites where no
observations were
made. Inserting additional unobserved sites can also be useful in
other cases. For instance, if $A^*$ cuts off a substantial part of any
$A_j$, that is, $A_j\cap A^*$ is much smaller than $A_j$ but not empty,
the areal prediction might be improved
by replacing $Y_t(\matt s_j)$ by the prediction of $Y_t$ at the center of
gravity of $A_j \cap A^*$, or if the area $A^*$ is small and
contains only a few stations, improved predictions of the areal average can
be obtained by making predictions at a few additional points inside the area.

\section{Application to short term prediction of precipitation}\label{SecApl}
We apply the model to obtain short term forecasts of precipitation.
Such forecasts are important, for instance, for agriculture and
flooding. The traditional way for obtaining precipitation forecasts is
the use
of numerical weather prediction (NWP) models. NWP models solve complex,
nonlinear equations emulating the dynamics of the atmosphere.
Typically, NWP models require a lot of computational resources to run.
Fitting our statistical model using the MCMC algorithm presented
above is also computationally intensive. However, once the statistical model
is fitted and assuming that the posterior of the primary parameters does
not change (see Section~\ref{AppPred} for more details), predictions
are computationally a lot
cheaper. Furthermore, the statistical model can be used in
situations where there are no NWP models available or to obtain
predictions at
different temporal resolutions than the one at which the NWP model operates.

\subsection{The data}\label{datadesc}

\begin{figure}

\includegraphics{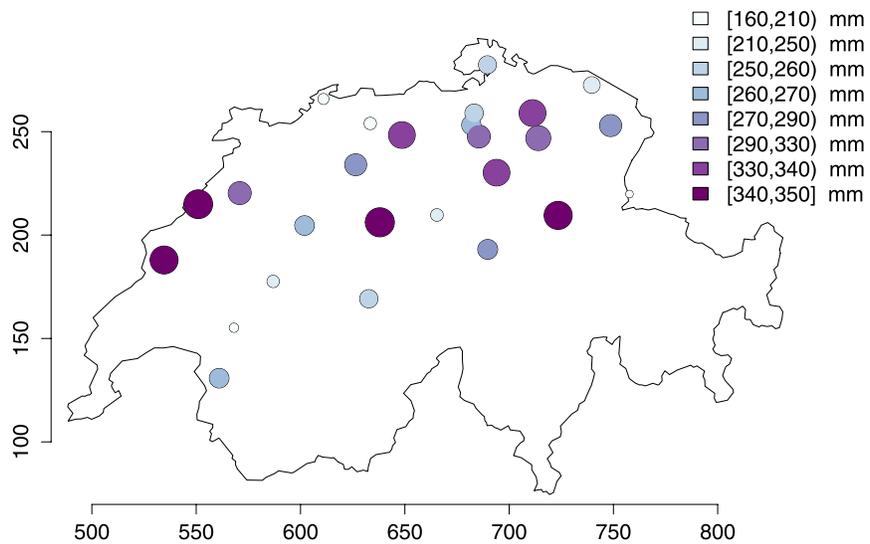}

\caption{Illustration
of the spatial distribution
of precipitation. The circles
display the cumulative rainfall amounts over time at the stations. The
larger the
circle and the darker the color, the higher is the cumulative precipitation
amount. Both axes are in km.}
\label{fig:RainVsSpace}
\end{figure}

The data consists of three-hourly precipitation amounts
collected by 26 stations around the Swiss
Plateau
from the beginning of December 2008 to the end of March 2009, making a
total of $968$ time periods. The data were
provided by MeteoSwiss. We use the
first three months, consisting of $720$ time periods, for fitting the
model. The remaining month March, consisting of $248$ time periods, is
%
%
\begin{figure}

\includegraphics{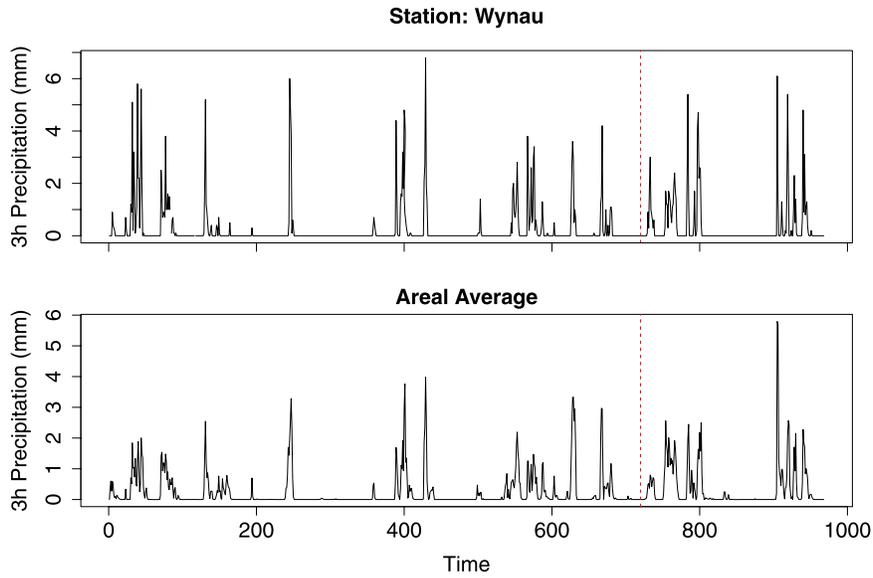}

\caption{Precipitation versus time. The lines are
observed precipitation of one station (corresponding to the station
with the acronym WYN in Figure \protect\ref{fig:Stations}) and of the
areal average. The time axis is in 3~h steps starting at December 1,
2008. The dotted vertical line separates the training and test data.}
\label{fig:RainVsTime}
\end{figure}
set aside for model evaluation. The locations of these
stations are shown in Figure~\ref{fig:Stations}. In Figure \ref
{fig:RainVsTime}, a time series plot
of the observed precipitation at one station (corresponding to the
station with the acronym WYN in Figure~\ref{fig:Stations}) and of the
weighted areal average is
shown. Concerning the latter, we take the weighted average over the
entire spatial domain. Figure~\ref{fig:RainVsSpace} shows the spatial
distribution
of the precipitation accumulated over time.

%
%

The covariates consist of the x- and y-coordinates (km), altitude (m),
temperature ($^{\circ}$C), dew point ($^{\circ}$C) and specific humidity
($\%$). Specific humidity is the ratio of water vapor to dry air in
a particular mass. It is expected to be positively related
to precipitation. The dew point is the temperature to which a given parcel
of humid air must be cooled, at constant barometric pressure, for water
vapor to condense into water. Thus, the lower the dew point, the lower
is the
chance for precipitation. However, specific humidity and dew point are
considerably negatively correlated. This makes it unclear, a priori, what
their joint relation to precipitation is like. Temperature, dew point
and specific humidity are predicted
variables obtained from an NWP model called COSMO-2. From the same
model, we also obtain
wind predictions (speed is in m/s). Predictions of the
statistical model are evaluated by comparing them to precipitation
forecasts from the same NWP. Having a high resolution with a grid
spacing of
2.2 km, the NWP model is able to resolve convective dynamics. The NWP
model produces predictions once a day
for 24 hours ahead starting at 0:00UTC. After assimilation and computation,
forecasts are available at around 1:30UTC. For all meteorological
variables, we use values at approximately
$1000$ m above ground. This is the
height where we think these variables to be most influential for
precipitation. All covariates are centered and standardized to unit
variance. Centering covariates around their means is used in
order to avoid correlations of the regression coefficients with
the intercept and to reduce posterior correlations.

\subsection{Fitting and results}\label{FitRes}
In the following, the nonstationary anisotropic model incorporating the
wind as an external drift term (see Section~\ref{SecMod}) is fitted. In
addition, we also fit a separable model. We simulate from the posterior
distributions of these models as outlined in
Section~\ref{SecFit}.

After the burn-in period consisting of $5000$ draws, $195\mbox{,}000$ samples
from the Markov chain were used to characterize posterior
distributions. Convergence was monitored by inspecting trace plots.


%
%
\begin{table}[b]
\caption{Posterior modes and 95\% credible intervals for the
nonstationary, anisotropic model with an external drift}
\label{PostQuant}
\begin{tabular*}{\tablewidth}{@{\extracolsep{\fill}}ld{2.6}d{2.6}d{2.5}@{}}
\hline
& \multicolumn{1}{c}{\textbf{Mode}} & \multicolumn{1}{c}{\textbf{2.5\%}}
& \multicolumn{1}{c@{}}{\textbf{97.5\%}} \\
\hline
Intercept & -1.05 & -1.21 & -0.929 \\
$X$ & -0.0473 & -0.133 & 0.0541 \\
$Y$ & -0.0108 & -0.0846 & 0.0531 \\
$Z$ & 0.00347 & -0.0169 & 0.0247 \\
Temp & -0.717 & -0.856 & -0.583 \\
Dew point & 0.406 & 0.187 & 0.601 \\
Spec hum & 1.14 & 0.949 & 1.33 \\
$\lambda$ & 1.58 & 1.54 & 1.62 \\
$\tau^2$ & 0.0685 & 0.0451 & 0.0943 \\
$\sigma^2$ & 1.04 & 0.953 & 1.17 \\
$\rho_0$ & 92 & 86.4 & 97.9 \\
$\phi$ & 0.000159 & 0.000147 & 0.00017 \\
$\rho_1$ & 93.6 & 88.1 & 99.4 \\
$c$ & 4.1 & 3.61 & 4.63 \\
$\alpha$ & 0.704 & 0.658 & 0.777 \\
$u$ & 0.879 & 0.645 & 1.1 \\
\hline
\end{tabular*}
\end{table}

In Table~\ref{PostQuant} we show posterior modes as well as 95\%
credible intervals for the different parameters of the nonstationary
anisotropic drift model. The coefficients of the geographic coordinates
are not significant. Specific humidity has a large positive
coefficient. As
expected, higher humidity implies more rainfall. The dew point is also
positively related to precipitation. Higher temperatures, on the other
hand, seem to imply less precipitation.

%
%
\begin{figure}

\includegraphics{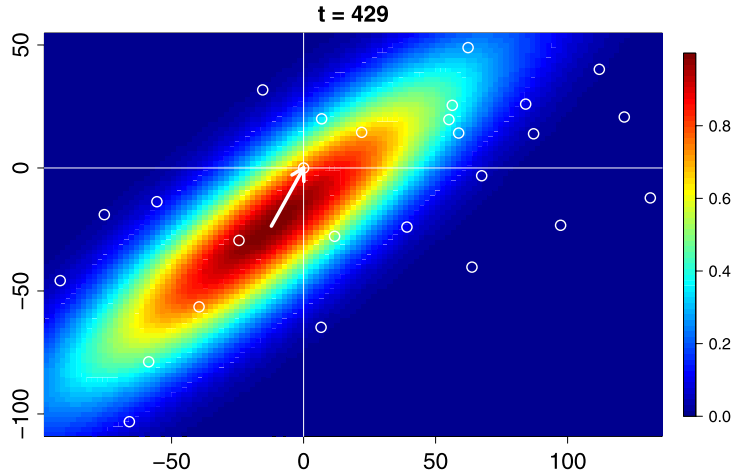}

\caption{Illustration of the convolution kernel at time $t=429$. The colors
indicate the lag-1 influence of the other stations on the station
Wynau. The white arrow represents the drift caused by a south-west
wind at this time point.
The dots represent the observation stations.
The axes are
in km.}
\label{fig:KernelIllus}
\end{figure}

For interpreting the fitted parameters governing the convolution kernel
($\rho_1$, $c$, $\alpha$ and $u$), we illustrate in Figure \ref
{fig:KernelIllus} the convolution kernel over the region where the
stations lie. The parameters $\rho_1$, $c$, $\alpha$ and $u$ are taken
at their posterior mode. The plot is interpreted as follows. The height
of the kernel is the level of influence that
$\vect{\xi}_{t-1}(\matt{s}')$ at location $\matt{s}'$ has on
$\vect{\xi}_{t}(\matt{s})$ at location $\matt{s}$
as a function of $\matt{s}'-\matt{s}$. In other words, the colors
represent the
lag-1 influence of the other stations on the station Wynau which is
used as
origin in the plot. The white arrow represents the drift vector
$\vect{\mu}_t=u\cdot\matt{w}_t$ at
time $t=429$, $\matt{w}_t$ being the wind vector. Note that this
transport vector changes over time, thus
causing temporal nonstationarity. The time $t=429$ illustrates a
meteorological situation with the typically
predominant southwestern wind direction.

With $c$ and $\alpha$ being approximately $4$ and $0.7$, we
observe anisotropy along the south-east north-west direction.
This corresponds to the topography of the region, as the area containing
a majority of the stations lies between two
mountain ranges: the Jura to the north-west and the Alps to the
south-east.
Correlations are expected to be higher along the flat part between
these two
mountain ranges.

Furthermore, the plot shows how the external drift shifts the convolution
kernel. Apparently, the southwestern neighbor (Bern) has the highest
influence on Wynau in this situation, with wind coming from the
southwest. \citet{GnEtAl06} observe a similar
phenomenon in wind speed data over the U.S. Pacific Northwest where
there is
also a predominant wind direction causing asymmetric cross-correlations.

\subsection{Short term prediction of precipitation}\label{AppPred}
In the following, we apply the fitted models to produce short term
predictions of precipitation. As mentioned before, we have fitted the model
to the first $720$ time periods from December 2008 to February 2009. From
this we obtain posterior distributions for the primary
parameters. Predictions for the time periods in March that were set
aside are
obtained as described in the following.

Ideally, one would run the full MCMC
algorithm at each time point, including all data up to the point,
and obtain predictive distributions from
this. However, since this is rather time consuming, we make the following
approximation. We assume that the posterior
distribution of the primary parameters given
$\matt{Y}_{1\dvtx t}=\{\matt{Y}_1,\ldots,\matt{Y}_{t}\}$ is the same
for all
$t\geq720$. That is, we neglect the additional information that the
observations in March give about the primary parameters. In practice,
this means
that posterior distributions of the primary parameters are calculated only
once, namely, on the data set from December 2008 to February 2009.

For each time $t \geq720 $, we make up to $8$ steps ahead forecasts.
That is, we sample from the
predictive distribution of $\matt{Y}^*_{t+k}$, $k= 1,\ldots,8$, given
$\matt{Y}_{1\dvtx t}=\{\matt{Y}_1,\ldots,\matt{Y}_{t}\}$
and given the posterior
of the primary parameters based on the data from December 2008 to February
2009. Since the NWP produces forecasts for the three meteorological
covariates once a day, for each prediction time $t+k$, the forecasts
made at 0:00UTC of
the same day are used. Sampling from the predictive distribution
consists of
imputing the augmented data $\matt W$ and sampling from the latent process
$\vect{\xi}$. These two steps are done as described in Section \ref
{SecFit}. To generate one sample from the predictive distribution
takes around 3.5 seconds on an AMD Athlon(tm) 64 X2 Dual Core Processor
5600$+$ with a 2900 MHz CPU clock rate. We use 200 samples to characterize
each predictive distribution.

The assumption that the posterior of the primary parameters does not
change may
be questionable over longer time periods and
when one moves away from the time period from which data is used to
obtain the posterior distribution. But
since all our data lies in the winter season, we think that this
assumption is
reasonable. If longer time periods are considered, one could use sliding
training windows or model the primary parameters as evolving dynamically
over time.
One can also investigate how the predictive performance
deteriorates with increasing lags between predictions and last time
point from
which data is used to fit the model.

In addition to the separable model and the nonstationary anisotropic
drift model, we fit a model with no autoregressive term, that is, with
$\phi=0$. Further, to assess\vadjust{\goodbreak}
how much information stems from the three meteorological
covariates (temperature, dew point and specific humidity) and how much
from the dynamic spatio-temporal model, we also fit the
nonstationary anisotropic drift model without including these
covariates. For each model, we calculate pointwise predictions for the
individual stations and also predictions for the areal average. The
latter are obtained using the Voronoi tessellation as described in Section
\ref{Prediction}.

In order to asses the performance of the probabilistic predictions, we
use the continuous
ranked probability score (CRPS) [\citet{MaWi76}]. The
CRPS is a strictly proper scoring rule [\citet{GnRa07}] that
assigns a
numerical value to probabilistic forecasts and assesses calibration and
sharpness simultaneously [\citet{GnBaRa07}]. It is defined as
%
%
\begin{equation}
\operatorname{CRPS}(F,y)=\int_{-\infty}^{\infty}\bigl(F(x)-
\mathbf{1}_{\{y \leq x \}}\bigr)^2\,dx,
\end{equation}
where $F$ is the predictive cumulative distribution function, $y$ is
the observed
realization, and $\mathbf{1}$ is an indicator function. It can be
equivalently calculated as
%
%
\begin{equation}
\operatorname{CRPS}(F,y)=E_F|Y-y|-\tfrac{1}{2}E_F\bigl|Y-Y'\bigr|,
\end{equation}
where $Y$ and $Y'$ are independent random variables with distribution
$F$. If a sample $Y^{(1)},\ldots, Y^{(m)}$ from $F$ is available, it
can be
approximated by
%
%
\begin{equation}
\frac{1}{m}\sum_{i=1}^m\bigl|Y^{(i)}-y\bigr|-
\frac{1}{2m^2}\sum_{i,j=1}^m\bigl|Y^{(i)}-Y^{(j)}\bigr|.
\end{equation}

\begin{figure}

\includegraphics{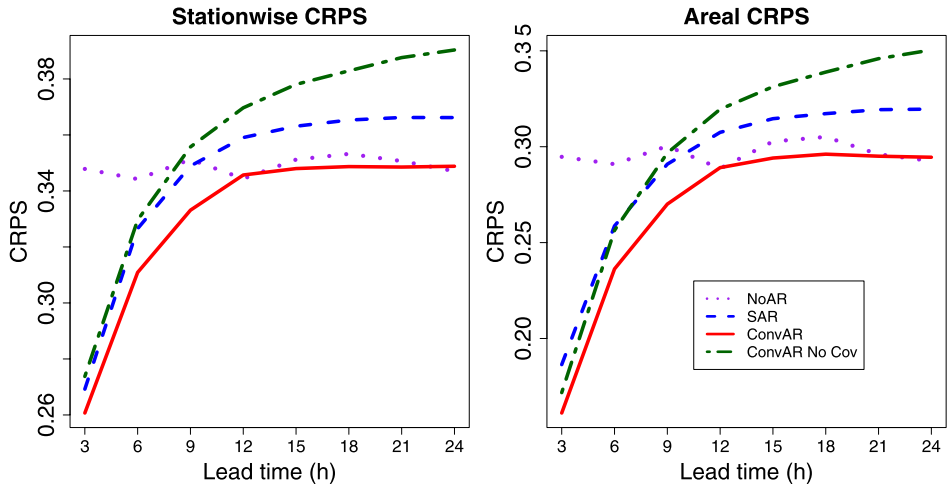}

\caption{Comparison of statistical models. The continuous ranked
probability score (CRPS) of forecasts versus
number of
consecutive time periods
for which predictions are made is shown. On the left are CRPSs of
station specific
forecasts and on the right are CRPSs of areal forecasts. ``NoAR'' denotes
the model without an autoregressive term, ``SAR'' the one with a separable
covariance structure, and ``ConvAR'' the convolution based nonstationary
anisotropic drift model. All three models include the covariates described
in Section \protect\ref{datadesc}. A convolution based model without including
covariates (``ConvAR No Cov'') is also fitted. The unit of the CRPS is mm.}
\label{fig:CRPS}
\end{figure}

In Figure~\ref{fig:CRPS} the average CRPS of the pointwise predictions and
the areal predictions are plotted versus lead times. In the left plot,
the mean is taken over all stations and
time periods, whereas the areal version is an average over all time
periods. Predictions $\matt{Y}^*_{t+k}$, $k= 1,\ldots,8$, for the next 8
time steps are made at each time point $t$.
We recall that the NWP model produces predictions for 8 consecutive
periods once a day at midnight. For simplicity, potential diurnal
variation in the accuracy of the predicted covariates is ignored.

%
%

We see that the nonstationary anisotropic drift model (``ConvAR'') has
clearly the best
performance among the three models. In particular, the nonseparable
convolution based model performs better than the simpler separable
spatio-temporal model (``SAR''). Not surprisingly, the model without
temporal dependency (``NoAR'') performs worse than the other two
models. Comparing the ``\mbox{ConvAR}'' model, the nonstationary convolution model
without covariates (``\mbox{ConvAR} No Cov''), and the ``NoAR'' model, we see
that the
main source of predictive performance at small lead times is not the
covariates but the
dynamic spatio-temporal model. In the areal case, the nonstationary
convolution model without covariates even outperforms the simple
autoregressive model including covariates at small
lead times. With increasing lead time, the meteorological covariates
contribute more to the predictive performance and the dynamic
spatio-temporal model becomes less important.

We also compare the performance of the predictions from the nonstationary
anisotropic drift model with predictions obtained from the
NWP model. Since the NWP model produces
deterministic forecasts, we use the mean absolute error (MAE). In order to
make the comparison fair, we first reduce the statistical
distributional forecast to a point forecast by taking the
median [see \citet{GnJASA11} on why this is a reasonable choice].
As mentioned,
the NWP model produces predictions once a day starting at
0:00UTC. Predictions are then made for eight consecutive
time periods corresponding to 24 h ahead. This means that the time of
day also
corresponds to the lead time. This is in contrast to
the above comparison of the different statistical models where 8 step
ahead predictions were made at all time periods.

\begin{figure}

\includegraphics{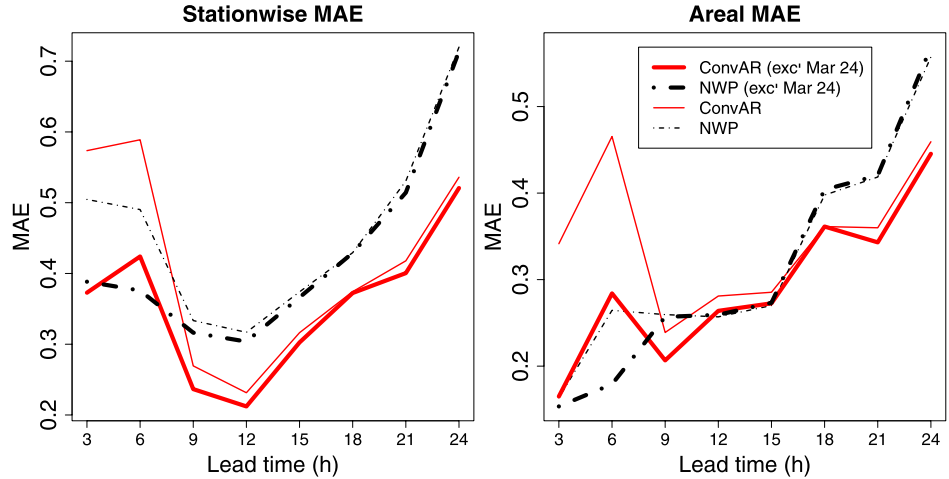}

\caption{Comparison of statistical and NWP model. The mean absolute error
(MAE) of forecasts versus lead time is shown. Lead time
also corresponds to the time of day. The left panel shows MAEs of
station specific
forecasts averaged over time and the stations, and the right panel shows
MAEs of areal forecasts averaged over time. ``ConvAR'' denotes the
convolution based nonstationary
anisotropic drift model and ``NWP'' the NWP model. The bold lines show the
results when excluding March 24, 2009. The unit of the MAE is mm.}
\label{fig:CRPSCosmo}
\end{figure}

In Figure
\ref{fig:CRPSCosmo} the mean absolute error
(MAE) of forecasts versus lead time, or time of day is
%
%
shown. In addition, in Table
\ref{fig:AverageMAE} we
report MAEs averaged over all lead times. Note that there is one
particular day (March 24)
when heavy rainfall occurred shortly after 0:00UTC. We report results
including (thin lines) and excluding (bold lines) this day.

%
%
\begin{table}[b]
\caption{Comparison of statistical and NWP model. The mean absolute error
(MAE) averaged~over all days and lead times is reported.
``ConvAR'' denotes the~convolution~based nonstationary
anisotropic drift model and\break ``NWP'' the NWP model. The unit of the MAE
is mm}\label{fig:AverageMAE}
\begin{tabular*}{\tablewidth}{@{\extracolsep{\fill}}lcccc@{}}
\hline
& \textbf{ConvAR} & \textbf{NWP} & \textbf{Areal ConvAR} & \textbf{Areal NWP} \\
\hline
March 2009 & 0.41 & 0.46 & 0.35 & 0.32 \\
Excluding March 24 & 0.36 & 0.43 & 0.29 & 0.31 \\
\hline
\end{tabular*}
\end{table}

Table~\ref{fig:AverageMAE} shows that overall the
statistical model outperforms the NWP on a stationwise base. When
considering the areal average, the two models perform similarly. Depending
on whether March 24 is included or not, the NWP or the statistical
model has a slightly lower average MAE.

Furthermore, Figure~\ref{fig:CRPSCosmo} shows that March 24
considerably affects the performance of the one- and two-step
ahead predictions of the statistical model as well as the
stationwise performance of the NWP model. When excluding this day, the
corresponding MAEs are considerably lower. This shows a typical
behavior of our model and statistical models in general: they perform
well when, at the time of prediction, the major phenomena
(advective fronts) are already observable. In
this case, the spatio-temporal statistical model can extrapolate the space--time
dynamics of the rainfall process into the future.

Earlier studies have shown that nowcasting methods, including
statistical approaches,
perform usually better at short lead times (up to one day), while NWP
have higher predictive
skills at medium ranges [see \citet{KoEtAl11} or \citet
{LittleEtAl09}]. Our
results are in line with these findings in the sense that all lead
times used in our application are still in
the range of what is considered ``short'' lead times. However, our
model is
not just based on past precipitation observations but also on other
predicted meteorological variables.

\section{Conclusions}\label{SecCon}
A hierarchical Bayesian spatio-temporal model is presented. Incorporating
physical knowledge, the dynamic model is nonstationary,
aniso\-tropic, and allows for nonseparable covariance structures. It
incorporates a drift term that depends on a wind vector. At the data
stage, the model determines the probability of rainfall and the
rainfall amount
distribution together. 
The model is fitted using
Markov chain Monte
Carlo (MCMC) methods and applied to obtain short term precipitation
forecasts. It performs better
than a separable, stationary and isotropic model, and it performs
comparably to a deterministic numerical weather prediction model and has
the advantage that it quantifies prediction uncertainty.

Even though we have applied the model to prediction of precipitation,
it can
also be used to predict or interpolate other meteorological quantities
of interest.

Future research could focus on adapting the model so that in can be applied
to spatially highly
resolved data. Using Markov random fields [\citet{RuHe05},
\citet
{LiLiRu10}] for the innovation process
$\vect{\varepsilon}_t$ might be a potential direction.
Alternatively, a dimension reduction approach could
be examined; cf.
\citet{BaEtAl08}. For instance, \citet{SiKuSt12}
approximate an
advection-diffusion SPDE to cope with large data sets. Further, the
model can be extended by additionally relaxing some
assumptions. For instance, the parameters $\sigma^2$, $\phi$, $\rho_0$,
$\rho_1$ and $\lambda$ were assumed to be constant over time. Assuming
periodicity, Fourier harmonics could be used to model parameters that vary
seasonally during the year. Alternatively, the parameters could evolve
dynamically over
time according to an equation of the form
$\vartheta_t=\vartheta_{t-1}+N(0,\sigma^2_{\vartheta})$.

\section*{Acknowledgments}

We thank Vanessa Stauch from
MeteoSwiss for providing parts of the data and for interesting
discussions. We also would like to thank the Editor and three anonymous
referees for their insightful comments and suggestions.


%
%

\printaddresses

\end{document}